# Structural, magnetic and transport properties in the Pr-doped manganites $La_{0.9-x}Pr_xTe_{0.1}MnO_3$ ( $0 \leq x \leq 0.9$ )

J.Yang [1], W. H. Song[1], Y. Q. Ma[1], R. L. Zhang[1], B. C. Zhao[1], Z.G. Sheng[1], G. H. Zheng[1], J. M. Dai[1], and Y. P. Sun*[1,2]

[1]*Key Laboratory of Materials Physics, Institute of Solid State Physics, Chinese Academy of Sciences, Hefei, 230031, P. R. China*

[2]*National Laboratory of Solid State Microstructures, Nanjing University, Nanjing 210008, P. R. China*

**Abstract**

The effect of Pr-doping on structural, magnetic and transport properties in electron-doped manganites $La_{0.9-x}Pr_xTe_{0.1}MnO_3$ $(0 \leq x \leq 0.9)$ with fixed carrier concentration are investigated. The room temperature structural transition from rhombohedral ( $R\bar{3}C$ ) to orthorhombic (*Pbnm*) symmetry is found in the samples with $x \geq 0.36$ by the Rietveld refinement of x-ray powder diffraction patterns. The Curie temperature $T_C$ of samples decreases and the transition becomes broader with increasing Pr-doping level. For the samples with $x \leq 0.36$, there exist insulator-metal (I-M) transition. And the low-temperature I-M transition is observed at about 66K for the sample with x = 0.36, which may be related to the opening of a new percolation channel. For the samples with $x \geq 0.54$, ρ(T) curves display the semiconducting behavior ($d\rho/dT < 0$) in both high-temperature PM phase and low-temperature FM phase. The results are discussed in terms of the increased bending of the Mn-O-Mn bond with decreasing the average ionic radius of the A-site element $<r_A>$ and the tolerance factor $t$, resulting in the narrowing of the bandwidth, the decrease of the mobility of $e_g$ electrons and the weakening of DE interaction caused by the substitution of smaller $Pr^{3+}$ ions for larger $La^{3+}$ ion.

## I. Introduction

Mixed-valent manganites perovskites have attracted considerable attention in recent years because of the observation of colossal magnetoresistance (CMR) and more generally due to the unusually strong coupling between their lattice, spin, and charge degrees of freedom. Although the focus of interest has primarily rested with the hole-doped manganites $Ln_{1-x}A_xMnO_3$ (Ln = La-Tb, and A = Ca, Sr, Ba, Pb, etc.) due to their potential applications such as magnetic reading heads, field sensors and memories,[1-3] naturally many researches have placed emphasis on electron-doped compounds such as $La_{1-x}Ce_xMnO_3$,[4-7] $La_{1-x}Zr_xMnO_3$,[8] $La_{2.3-x}Y_xCa_{0.7}Mn_2O_7$,[9] and $La_{1-x}Te_xMnO_3$[10-12] because having both electron as well as hole doped ferromagnetic (FM) manganites may open up very interesting applications in the emerging field of spintronics. These investigations also suggest that the CMR behavior probably occur in the mixed-valent state of $Mn^{2+}/Mn^{3+}$. The basic physics in terms of Hund's rule coupling between $e_g$ electrons and $t_{2g}$ core electrons and Jahn-Teller (JT) effect due to $Mn^{3+}$ JT ions can operate in the electron-doped manganites as well.

It is well known that for hole-doped manganites, the following two factors have been shown to mainly affect the DE interaction, i.e., the hole carriers density controlled by the $Mn^{3+}/Mn^{4+}$ ratio and the average ionic radius of the A-site element $<r_A>$.[13-19] From the point of view of being favorable to stabilize the low-temperature FM metallic phase one would expect an optimum $Mn^{3+}/Mn^{4+}$ ratio to be 2:1. On the other hand, the optimum $Mn^{3+}/Mn^{4+}$ ratio is favorable to form an ideal cubic perovskite. Any deviation from the ideal cubic perovskite would lead to a reduction in the Mn-O-Mn bond angle from 180°, which directly weakens the DE. Beside the $Mn^{3+}/Mn^{4+}$ ratio, the average ionic radius of the A-site element $<r_A>$ has also been shown to influence the DE. The principal effect of decreasing $<r_A>$ is to reduce the Mn-O-Mn bond angle, thereby reducing the matrix element b that described electron hopping between Mn sites, which is confirmed by Hwang et al.[14] and Fonteuberta et al.[15] A salient question to ask is: what is the case in electron-doped manganites? With this notion in mind, we have examined a series of samples in which the average ionic radius of the A-site element $<r_A>$ is systematically varied while keeping the $Mn^{2+}/Mn^{3+}$ ratio fixed at 1/9. We find that the average ionic radius of the A-site elemen $<r_A>$ has strongly affected the structural, magnetic and transport properties in electron-doped manganites samples $La_{0.9-x}Pr_xTe_{0.1}MnO_3$ $(0 \leq x \leq 0.9)$.

## II. Experiment

A series of ceramic samples of $La_{0.9-x}Pr_xTe_{0.1}MnO_3$ $(0 \leq x \leq 0.9)$ were synthesized by a conventional solid-state reaction method in air. The powders mixed in stoichiometric compositions of high-purity $La_2O_3$, $Pr_6O_{11}$, $TeO_2$ and $MnO_2$ were ground, then fired in air at 700°C for 24h. The powders obtained were ground, pelletized, and sintered at 1050°C for 24h with three intermediate grindings, and finally, the furnace was cooled down to room temperature. The structure and lattice constant were determined by powder x-ray diffraction (XRD) using $CuK_\alpha$ radiation at room temperature. The resistance as a function of temperature was measured by the standard four-probe method from 25 to 300K. The magnetic measurements were performed on a Quantum Design superconducting quantum interference device (SQUID) MPMS system (2 ≤ T ≤400 K, 0 ≤ H ≤5 T).

## III. Results and discussion

X-ray powder diffraction (XRD) at room temperature shows that all samples are single phase with no detectable secondary phases. XRD patterns of the samples with x = 0 and x = 0.18 can be indexed by rhombohedral lattice with space group $R\bar{3}C$. While XRD patterns of the samples with x = 0.36, 0.54, 0.72 and 0.9 can be indexed by orthorhombic lattice with space group $Pbnm$. The structural parameters are refined by the standard Rietveld technique [20] and the fitting between the experimental spectra and the calculated values is relatively good based on the consideration of lower $R_P$ values as shown in Table I. Figs. 1(a) and 1(b) show experimental and calculated XRD patterns for the samples with x = 0 and 0.36, respectively. The structural parameters obtained are listed in Table I. As we can see, for samples $La_{0.9-x}Pr_xTe_{0.1}MnO_3$ $(0 \leq x \leq 0.9)$, the crystal structure at room temperature changes from rhombohedral phase ($R\bar{3}C$, Z = 2, $x \leq 0.18$) to orthorhombic phase (*Pbnm*, Z = 4, $x \geq 0.36$). The lattice distortion and the bend of Mn-O-Mn bond increase when the crystal structure varies from rhombohedral lattice to orthorhombic lattice. It is well known that one of the possible origins of the lattice distortion of perovskites structures is the deformation of the $MnO_6$ octahedra originating from JT effect that is directly related to the concentration $Mn^{3+}$ ions. But for the present study samples, the concentration of $Mn^{3+}$ ions is fixed. And thus the observed lattice distortion should be only caused by the average ionic radius of the A-site element $<r_A>$, which is governed by the tolerance factor $t$ [$t = (r_A + r_O)/\sqrt{2}(r_B + r_O)$], where $r_i$ ($i$=A, B, or O) represents the average ionic size of each element. As $t$ is close to 1, the cubic perovskite structure is expected to form. As $<r_A>$ decreases, so does $t$, the lattice structure transforms to rhombohedral ($R\bar{3}C$), and then to orthorhombic (*Pbnm*) structure, in which the bending of the B-O-B bond increases and the bond angle deviates from 180°. For $La_{0.9-x}Pr_xTe_{0.1}MnO_3$ samples, the structural transition at room temperature mainly originates form the variation of the tolerance factor $t$ induced by the substitution of smaller $Pr^{3+}$ for larger $La^{3+}$ ions.

Fig.2 shows the temperature dependence of magnetization M of $La_{0.9-x}Pr_xTe_{0.1}MnO_3$ $(0 \leq x \leq 0.9)$ under both zero-field cooling (ZFC) and field cooling (FC) modes at H = 0.1 T. The Curie temperature $T_C$ (defined as the one corresponding to the peak of $dM/dT$ in the M vs. T curve) are 239 K, 207 K, 159 K, 120K, 93K and 75K for x = 0, 0.18, 0.36, 0.54, 0.72 and 0.9, respectively, which are listed in Table II. Obviously, the Curie temperature $T_C$ decreases monotonically with increasing Pr-doping level. We suggest that the $T_C$ reduction should be attributed to the reduction of Mn-O-Mn bond angle with decreasing the average ionic radius of the A-site element $<r_A>$, and thereby reducing the matrix element $b$ which described electron hopping between Mn sites. Thus the DE interaction between $Mn^{2+}$-O-$Mn^{3+}$ becomes weakening because of the narrowing of the bandwidth and the decrease of the mobility of $e_g$ electrons due to the increase of Mn-O bond length and the decrease of Mn-O-Mn bond angle caused by the substitution of smaller $Pr^{3+}$ ions for larger $La^{3+}$ ions.

In addition, from Fig.2, a sharp FM to paramagnetic (PM) transition is observed for the samples with $x \leq 0.36$. However, as $x \geq 0.54$, the temperature range of FM-PM phase transition become broader with increasing Pr-doping level implying a wider distribution of the magnetic exchange interactions in the Mn-O-Mn network, i.e., the increase of magnetic inhomogeneity. Moreover, It is clear that the ZFC curve does not coincide with the FC curve below a freezing temperature $T_f$ for the samples with $x \geq 0.36$. With the increase of Pr-doping content, the difference between M-T curves under FC and ZFC modes becomes greater because of the increase of the magnetic frustration arising from the bending Mn-O-Mn bond, which is in accordance with the structural refinement results. This discrepancy between ZFC and FC magnetization is a characteristic of cluster glass.

The magnetization as a function of the applied magnetic field at 5K is shown in Fig.3. It shows that, for the samples with $x \leq 0.18$, the magnetization reaches saturation at about 1T and keeps constant up to 5T, which is considered as a result of the rotation of the magnetic domain under the action of applied magnetic field. For the sample with x = 0.36, the magnetization slowly reaches saturation at about 4T, implying the appearance of a small amount of AFM phase at low temperatures. However, for the samples with $x \geq 0.54$, Fig.3 exhibits that the rapid increase of magnetization M (H) at low magnetic fields resembles of ferromagnet with a long-range FM ordering corresponding to the rotation of magnetic domains, whereas the magnetization M increases continuously without saturation at higher fields, revealing a superposition of both FM and AFM components. The coexistence of and competition between ferromagnetic and antiferromagnetic interaction would favor the formation of a cluster glass state, as observed in $La_{0.9-x}Pr_xTe_{0.1}MnO_3$ ($x \geq 0.36$) samples. In fact, based on the temperature and magnetic field dependence of magnetization for these samples, the microscopic magnetic structure can be understood by presence of small sized FM clusters in the samples, as can be clearly observed from the broad magnetic transition range for the sample with x=0.9. Moreover, in order to determine the change in volume of the FM phase in respect to Pr doping level, a liner extrapolation of M (H) to H = 0 for the samples with $x \geq 0.54$ is plotted in dashed line in Fig.3. At 5K, the FM phase of the samples with x = 0.72 and 0.9 decreases by about 23% and 48%, respectively, in volume compared with that of the sample with x = 0.54. So it can be concluded that Pr-doping induces an increasing AFM superexchange interaction.

Fig.4 shows the temperature dependence of the inverse magnetic susceptibility $\chi_m$ for all samples. For ferromagnet, it is well known that in the PM region, the relation between $\chi_m$ and the temperature T should follow the Curie-Weiss law, i.e., $\chi_m = C/(T-\Theta)$, where $C$ is the Curie constant, and $\Theta$ is the Weiss temperature. The lines in Fig.4 are the calculated curves deduced from the Curie-Weiss equation. It can be seen from Fig.4 that the experimental curve in the whole PM temperature range is well described by the Curie-Weiss law. The Weiss temperature $\Theta$ is obtained to be 241K, 210K, 171K, 130K, 121K and 93K for the samples with x=0, 0.18, 0.36, 0.54, 0.72 and 0.90, respectively. For the samples with x = 0 and 0.18, $\Theta$ values almost approach their corresponding $T_C$ values. However, for the samples with x = 0.36, 0.54, 0.72 and 0.9, $\Theta$ values are higher than corresponding $T_C$ values which may be related to the magnetic inhomgeneity. The Curie constant C deduced from the fitting data is 6.05, 6.22, 6.45, 5.23, 5.21 and 4.40K·cm$^3$/mol for the samples with x = 0, 0.18, 0.36, 0.54, 0.72 and 0.90, respectively. And thus the effective magnetic moment $\mu_{eff}$ can be obtained as 4.923, 4.988, 5.082, 4.605, 4.566 and 4.198 $\mu_B$ for the samples with x = 0, 0.18, 0.36, 0.54, 0.72 and 0.90, respectively. According to a mean field approximation,[21] the expected effective magnetic moment $\mu_{eff}$ can also be calculated as $5\mu_B$ for the sample with x = 0, which is in accordance with x = 0.18 and 0.36 samples relatively well. The effective magnetic moment as a function of the Pr-doping level is shown in the inset of Fig.4. It indicates that the experimental $\mu_{eff}$ value increases with increasing Pr-doping content and there exists a maximum value for the sample with x = 0.36, and then the $\mu_{eff}$ value begins to decrease with further increasing Pr-doping level, which is mainly related to the occurrence of the structural transition. It is worth noting that the maximum $\mu_{eff}$ value appears in the sample with x = 0.36, just corresponding with the sample that occurs structural phase transition. In addition, for the sample with x = 0.36, the experimental $\mu_{eff}$ value is slightly higher than the expected $\mu_{eff}$ value. This phenomenon is also observed in $La_{1-\delta}MnO_3$, which is considered as the signature of clusters of $Mn^{4+}$ and $Mn^{3+}$.[21] Here, we ascribe it to larger magnetic moment of Pr ions. In addition, the fluctuating valence of $Pr^{3+}/Pr^{4+}$ is also a possible reason.

Fig.5 (a) shows the temperature dependence of resistivity ρ(T) for the samples with x = 0, 0.18 and 0.36 at zero fields in the temperature range of 30-300K. For sample with x = 0, it shows that there exists an insulator-metal (I-M) transition at $T_{P1}$ (= 246 K) which is close to its Curie temperature $T_C$ (= 239K). In addition, there exists a bump shoulder at $T_{P2}$ (= 223 K) below $T_{P1}$, which is similar to the double peak behavior observed usually in alkaline-earth-metal-doped and alkali-metal-doped samples of $LaMnO_3$.[22-26] More interesting phenomenon is that double I-M transitions show significant variation with changing the Pr-doping level. Double peaks ($T_{P1}$ = 210 and $T_{P2}$ = 186 K) shift to low temperatures for x = 0.18 sample. Compared with the x = 0 sample, I-M transition at $T_{P1}$ becomes weak and I-M transition at $T_{P2}$ becomes more obvious. It shows that the Pr-doping at La-site can substantially enhance the I-M transition at $T_{P2}$. When Pr-doping level is increased to x = 0.36, I-M transition at $T_{P1}$ (=153K) is almost invisible and displays an inflexion behavior, as can be seen from the Ln(ρ) vs. T curve. And I-M transition at $T_{P2}$ (=105K) becomes more obvious. In other words, the I-M transition at $T_{P1}$ has been almost suppressed. Moreover, there exists a low-temperature I-M transition at $T^*$ (= 66K) for the sample with x = 0.36, implying the presence of magnetic inhomogeniety due to the Pr-doping at La-site. Its real origin will be further explained below. The experimental data measured at applied field of 0.5 T for samples with x = 0, 0.18 and 0.36 in the temperature range of 30-300K are also recorded. It can be seen from Fig.5, for the samples with x = 0 and 0.18, the applied field suppressed the resistivity peak at $T_{P1}$ significantly and the resistivity peak shifts towards higher temperatures. Especially for the sample with 0.36, the resistivity peak at $T_{P1}$ seems to be suppressed completely under the applied field. However, for the second I-M transition at $T_{P2}$, it is worth noting that for the sample with x = 0 and 0.18, the applied field change the position of the resistivity peak at $T_{P2}$ slightly, whereas for the sample with x = 0.36 the position of the resistivity peak at $T_{P2}$ moves to higher temperature greatly under the applied magnetic field. The difference in the response of the resistivity peak at $T_{P2}$ for the applied field in samples between x = 0 and x = 0.18, 0.36 indicate that they may have different origins. As it can also be seen from Fig.6, for sample with x = 0 and 0.18, there exist corresponding peaks in the vicinity of $T_{P1}$ and a small hump at $T_{P2}$ on the magnetoresistance (MR) curves, which is similar to the MR behavior observed in polycrystalline $La_{1-x}Sr_xMnO_3$ samples.[27] Whereas for samples with 0.36, there exist one corresponding peak in the vicinity of $T_{P1}$ in the MR curve and the corresponding peak or hump at $T_{P2}$ is not observed although the applied field changes the position of the resistivity peak at $T_{P2}$ greatly. Moreover, for sample with x = 0.36, the MR curve also displays a low-temperature peak at about 65K corresponding to the temperature $T^*$. Here the MR is defined as $\Delta\rho/\rho_0 = (\rho_0 - \rho_H)/\rho_0$, where $\rho_0$ is the resistivity at zero field and $\rho_H$ is the resistivity at H = 0.5T. Additionally, the samples with x = 0, 0.18 and 0.36 all have evident MR at low temperatures, similar to the MR behavior observed usually in polycrystalline samples of hole-doped manganites, which is considered to be related to spin-dependent scattering at grain boundaries.[24, 28] So we consider the reason that the corresponding peak or hump at $T_{P2}$ for the sample with x = 0.36 does not appear arises mainly from the MR value near the temperature $T_{P2}$ being small compared with the large low-temperature MR. And thus the corresponding peak or hump at $T_{P2}$ for the sample with x = 0.36 is probably suppressed completely by the gradually ascending low-temperature MR.

As to the origin of the low-temperature I-M transition at about 66K for sample with x = 0.36, we consider it is mainly related to the opening of a new percolative channel. From Fig.5 (a), one can see that the ρ(T) curve under zero fields exhibits an upturn from 70K with further cooling, which is indeed the result of the competition between the AFM interaction and the ferromagnetic DE interaction in the sample. With further cooling, the amounts of small sized FM clusters increase and finally come into being a filament percolative channel. And thus the I-M transition at about 66K can be observed. Moreover, the applied field of 0.5T makes the temperature of percolation transition shift towards higher temperatures, as evidenced by presence of the corresponding MR peak at the temperature $T^*$ in the MR curve of the x=0.36 sample.

For the samples with x = 0.54, 0.72 and 0.9, ρ(T) curves display the semiconducting behavior ($d\rho/dT < 0$) in both high-temperature PM phase and low-temperature FM phase and the resistivity maximum increases by six orders of magnitude compared with that of no-Pr-doping sample implying the enhancement of the localization of carriers. This FM insulating (FMI) behavior is also found in $La_{1-x}Sr_xMnO_3$ [29, 30] and $La_{1-x}Li_xMnO_3$ [25] compounds with orthorhombic structure. FMI behavior cannot be explained based only on the DE model since the model requires the coexistence of the FM and metallic nature simultaneously. The FM order at low temperatures for the samples with x = 0.54, 0.72 and 0.9 can be understood by presence of small sized FM clusters. On the one hand, based on the coexistence of FM clusters and AFM insulating regions in low temperatures, the resistivity of the samples may be contributed to mainly from these insulating regions at low temperatures since the metallic clusters cannot develop into a whole network. On the other hand, the $La_{0.9-x}Pr_xTe_{0.1}MnO_3$ is a system with severe electrical and magnetic disorder due to the substitution of smaller $Pr^{3+}$ for larger $La^{3+}$ ions. The disorder may lead to electron localization and give rise to the high resistivity at low temperatures. Furthermore, the localization of $e_g$ electrons due to the structural transition from rhombohedral ($R\bar{3}C$) to orthorhombic (*Pbnm*) is also a reason. It is well known that the orthorhombic structure with space group *Pbnm* allows three independent Mn-O bonds as shown in Table I, therefore, it can accommodate a static coherent JT distortion of the $MnO_6$ octahedra, which provides an additional charge carrier localization.[31] Moreover, the change of the Mn-O-Mn bond angle has a substantial effect on the electronic transport due to the change of the bandwidth of $e_g$ electron. Usually the bandwidth of $e_g$ electron becomes narrow with the decrease of the θ value, which results in a charge-transfer insulator. So it can be concluded that FMI behavior in the orthorhombic samples with x = 0.54, 0.72 and 0.9 may arise mainly from the combined effects of the presence of FM clusters in low temperatures and the localization of $e_g$ electron caused by the disorder due to the Pr-doping at La-site and the structural transition.

The resistivity above $T_{P1}$ (corresponding to $T_C$) for the samples with x = 0, 0.18 and 0.36 fitted by variable range hopping model (VRH) [32] [$\rho \sim \exp(T_0/T)^{1/4}$] is shown in Fig.7 (a). The results show that ρ(T) curves can be well described by VRH model. Whereas for the samples with x = 0.54, 0.72 and 0.9, the results show that ρ(T) curves can be well fitted according to VRH model in the whole measurement temperature with two different fitting parameters of $(T_0)_1$ and $(T_0)_2$ in the different temperature range divided by the temperature denoted in the plot as shown in Fig.7 (b). The fitting parameter $T_0$, which is a characteristic temperature related to the localization length $\xi$ and the density of states $N(E_F)$ in the vicinity of Fermi energy level, i.e., $k_B T_0 \approx 21/[\xi^3 N(E_F)]$, is shown in Table II. From the Table II, it is found that the $T_0$ value increases obviously with the increase of Pr content, implying the decrease of the localization length and the reduction of the carrier mobility, which is intimately related to the localization of carriers and the destruction of DE interaction arising from Pr-doping at La-site, which is in accordance with the magnetic and electronic transport properties for the study samples.

Based on the above results, it can be concluded that the tolerance factor $t$ is the principal factor that strongly influence the structural, magnetic and transport properties in electron-doped manganites samples $La_{0.9-x}Pr_xTe_{0.1}MnO_3$ ($0 \leq x \leq 0.9$) because the carrier concentration is fixed. Standard ionic radii [33] for different element are used to calculate $t$ and $<r_A>$. The temperature phase diagram as a function of the tolerance factor $t$ and the average ionic radius of the A-site element $<r_A>$ is plotted in Fig.8. As we can see, with the decrease of $t$ and $<r_A>$, the Curie temperature $T_C$ of the study samples decreases as well as. It is worth noting that $T_C$ shows a linear dependence upon the tolerance factor $t$. Similar relation between the average ionic radius of the A-site element $<r_A>$ and $T_C$ is also observed. As $<r_A>$ decreases, so does $t$, the lattice structure transforms to rhombohedral ($R\overline{3}C$), and then to orthorhombic ($Pbnm$) structure. At the same time, the phase transition also occurs from PM-FMM to PM-FMI. All these are ascribed to the increase of the bending of the Mn-O-Mn bond with decreasing the average ionic radius of the A-site element $<r_A>$ and the reduction of the tolerance factor $t$ because of the substitution of smaller $Pr^{3+}$ ions for larger $La^{3+}$ ion.

## IV. Conclusion

The effect of Pr-doping on structural, magnetic and transport properties in electron-doped manganites $La_{0.9-x}Pr_xTe_{0.1}MnO_3$ $(0 \leq x \leq 0.9)$ with fixed carrier concentration has been studied systematically. The room temperature structure transition from rhombohedral ($R\bar{3}C$) to orthorhombic (*Pbnm*) symmetry is observed for the sample with $x \geq 0.36$. All samples undergo PM-FM phase transition and the Curie temperature $T_C$ of samples decreases with increasing the Pr-doping level. The high temperature peak in double-peak-like ρ(T) curves observed in no Pr-doping sample is almost suppressed as Pr-doping level x = 0.36. At the same time, there appear a new peak in the ρ(T) curve of the sample x = 0.36 at 66 K, which may be originated from the opening of a new percolation channel. For the samples with $x \geq 0.54$, ρ(T) curves display the semiconducting behavior ($d\rho/dT < 0$) in both high-temperature PM phase and low-temperature FM phase, which is considered to be related to the combined effects of the presence of FM clusters in low temperatures and the localization of $e_g$ electron caused by the disorder due to the Pr-doping at La-site and the structural transition.

**ACKNOWLEDGMENTS**

This work was supported by the National Key Research under contract No.001CB610604, and the National Nature Science Foundation of China under contract No.10174085, Anhui Province NSF Grant No.03046201 and the Fundamental Bureau, Chinese Academy of Sciences.

# Tables

**TABLE I.** Refined structural parameters of $La_{0.9-x}Pr_xTe_{0.1}MnO_3$ $(0 \leq x \leq 0.9)$ at room temperature. O(1):apical oxygen; O(2): basal plane oxygen.

| Parameter | x=0 | x=0.18 | x=0.36 | x=0.54 | x=0.72 | x=0.9 |
|---|---|---|---|---|---|---|
| a (Å) | 5.5241 | 5.5326 | 5.5156 | 5.5179 | 5.5180 | 5.5178 |
| b (Å) | 5.5241 | 5.5326 | 5.4865 | 5.4937 | 5.4935 | 5.4939 |
| c (Å) | 13.3572 | 13.3694 | 7.7811 | 7.7838 | 7.7839 | 7.7851 |
| v (Å$^3$) | 353.0103 | 354.4132 | 235.4668 | 235.9511 | 235.9534 | 235.9612 |
| Mn-O(1) (Å) | … | … | 1.9718 | 1.9722 | 1.9990 | 1.9998 |
| Mn-O(2) (Å) | … | … | 1.9906 | 2.0701 | 2.1157 | 2.1167 |
| Mn-O(2) (Å) | … | … | 1.9447 | 1.9219 | 1.8956 | 1.8941 |
| (Mn-O) (Å) | 1.9644 | 1.9718 | 1.9724 | 1.9880 | 2.0034 | 2.0035 |
| Mn-O(1)-Mn (º) | … | … | 161.18 | 157.23 | 156.42 | 156.27 |
| Mn-O(2)-Mn (º) | … | … | 162.55 | 161.62 | 161.59 | 161.43 |
| <Mn-O-Mn> (º) | 163.83 | 162.63 | 162.09 | 160.16 | 159.87 | 159.71 |
| R$_p$ (%) | 8.21 | 8.43 | 9.42 | 9.33 | 9.51 | 9.89 |

**TABLE II.** $T_C$, $T_P$ and the fitting parameter of $La_{0.9-x}Pr_xTe_{0.1}MnO_3$ $(0 \leq x \leq 0.9)$ samples.

| Parameter | x=0 | x=0.18 | x=0.36 | x=0.54 | x=0.72 | x=0.9 |
|---|---|---|---|---|---|---|
| $T_C$ (K) | 239 | 207 | 159 | 120 | 93 | 75 |
| $T_{P1}$ (K) | 246 | 210 | 153 | … | … | … |
| $T_{P2}$ (K) | 223 | 186 | 105 | … | … | … |
| $(T_0)_1$ | $2.36\times10^7$ | $8.65\times10^7$ | $9.38\times10^7$ | $1.97\times10^8$ | $2.08\times10^8$ | $3.09\times10^8$ |
| $(T_0)_2$ | … | … | … | $2.04\times10^8$ | $6.39\times10^8$ | $1.17\times10^9$ |

# Figure captions

Fig.1. XRD patterns of the compound $La_{0.9-x}Pr_xTe_{0.1}MnO_3$, (a) x = 0 and (b) x = 0.36. Crosses indicate the experimental data and the calculated data is the continuous line overlapping them. The lowest curve shows the difference between experimental and calculated patterns. The vertical bars indicate the expected reflection positions.

Fig.2. Magnetization as a function of temperature for $La_{0.9-x}Pr_xTe_{0.1}MnO_3$ (x = 0, 0.18, 0.36, 0.54, 0.72 and 0.9) measured at H = 0.1T under the field-cooled (FC) and zero-field-cooled (ZFC) modes that are denoted as the filled and open symbols, respectively.

Fig.3. Field dependence of the magnetization in $La_{0.9-x}Pr_xTe_{0.1}MnO_3$ (x = 0, 0.18, 0.36, 0.54, 0.72 and 0.9) at 5 K. The dashed lines represent the extrapolation lines and $M_0$ denotes a linear extrapolation M (H) to H = 0.

Fig.4. The temperature dependence of the inverse of the magnetic susceptibility for $La_{0.9-x}Pr_xTe_{0.1}MnO_3$ (x = 0, 0.18, 0.36, 0.54, 0.72 and 0.9) samples. The lines are the calculated curves according to the Curie-Weiss law. The inset is the variation of the effective magnetic moment with x and the dashed lines denote the boundaries between the different crystal structure symmetry.

Fig.5. (a)The temperature dependence of the resistivity of $La_{0.9-x}Pr_xTe_{0.1}MnO_3$ (x = 0, 0.18, 0.36) samples at zero (solid lines) and 0.5T fields (dashed lines). (b) The temperature dependence of the resistivity of $La_{0.9-x}Pr_xTe_{0.1}MnO_3$ (x = 0.54, 0.72, 0.9) samples at zero fields.

Fig.6. The temperature dependence of magnetoresistance (MR) ratio of $La_{0.9-x}Pr_xTe_{0.1}MnO_3$ at 0.5 T field for the samples with x = 0, 0.18 and 0.36.

Fig.7. The fitting plot of $\rho(T)$ curves of $La_{0.9-x}Pr_xTe_{0.1}MnO_3$ with x = 0, 0.18, 0.36, (a) and with 0.54, 0.72 and 0.9, (b) according to VRH model. The dashed lines represent the experimental data.

Fig.8. Phase diagram of temperature vs. tolerance factor t and the average ionic radius of the A-site element $<r_A>$ for $La_{0.9-x}Pr_xTe_{0.1}MnO_3$ (x = 0, 0.18, 0.36, 0.54, 0.72 and 0.9) samples. The mark PMI, FMM and FMI represent paramagnetic insulator, ferromagnetic metallic and ferromagnetic insulator phase, respectively. The dashed line denotes the boundaries between FMM and FMI

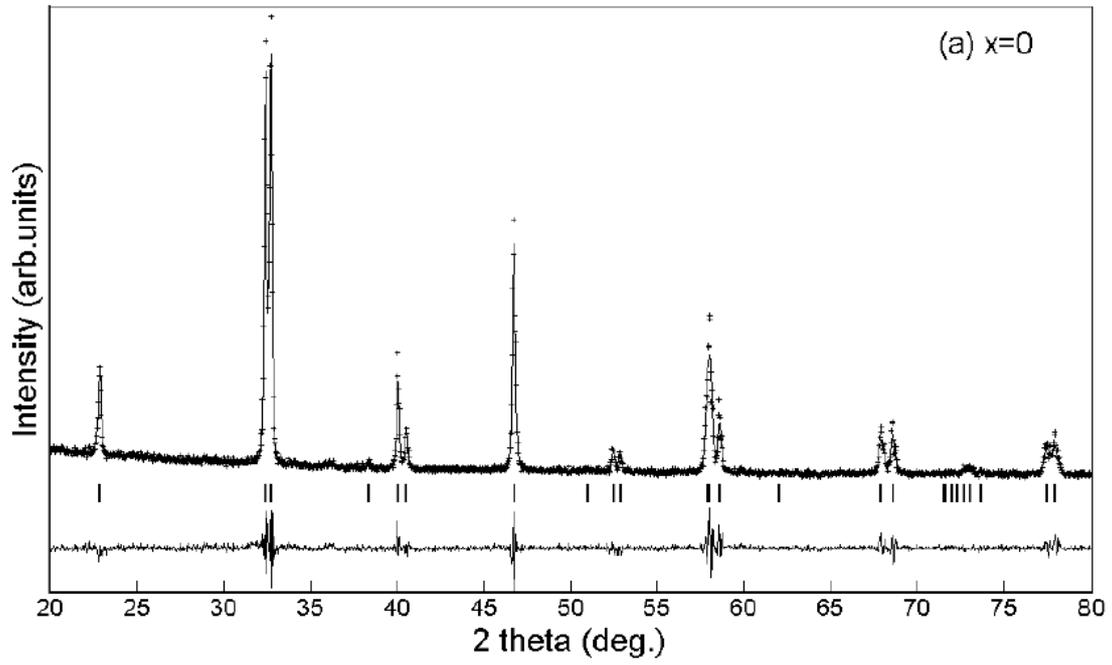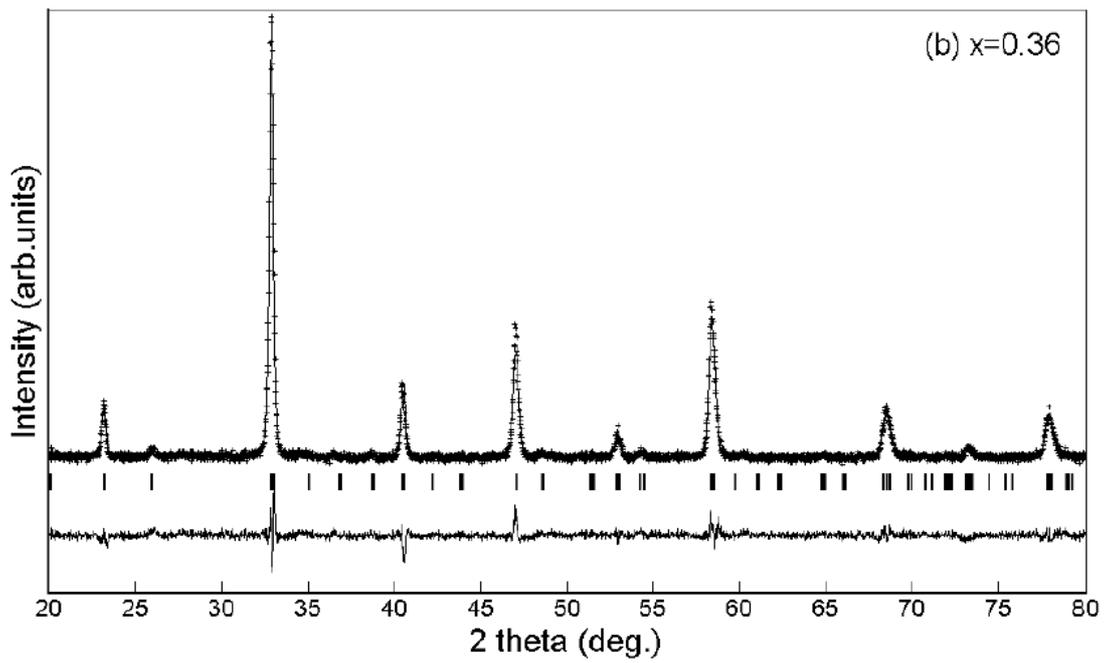

Fig.1 J. Yang et al.

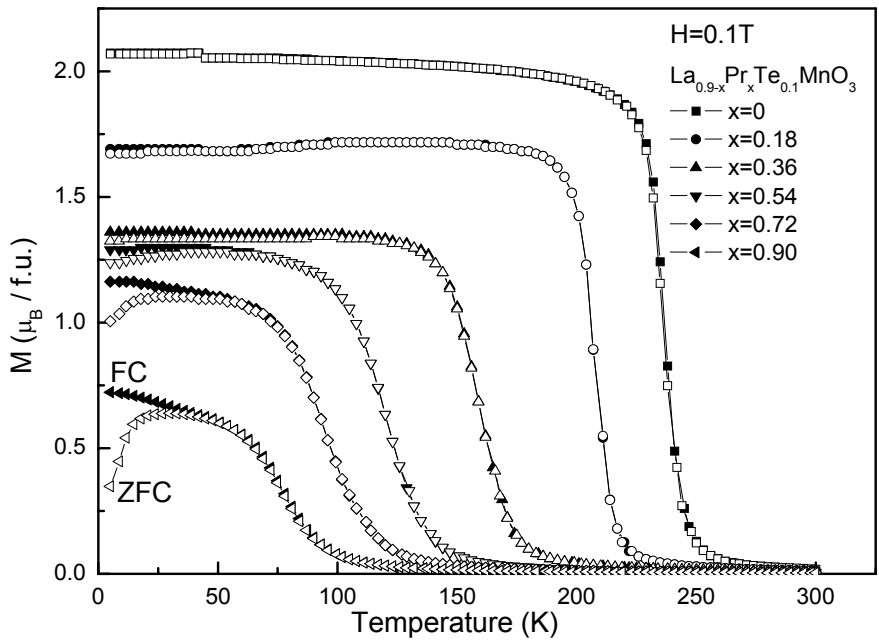

Fig.2 J. Yang et al.

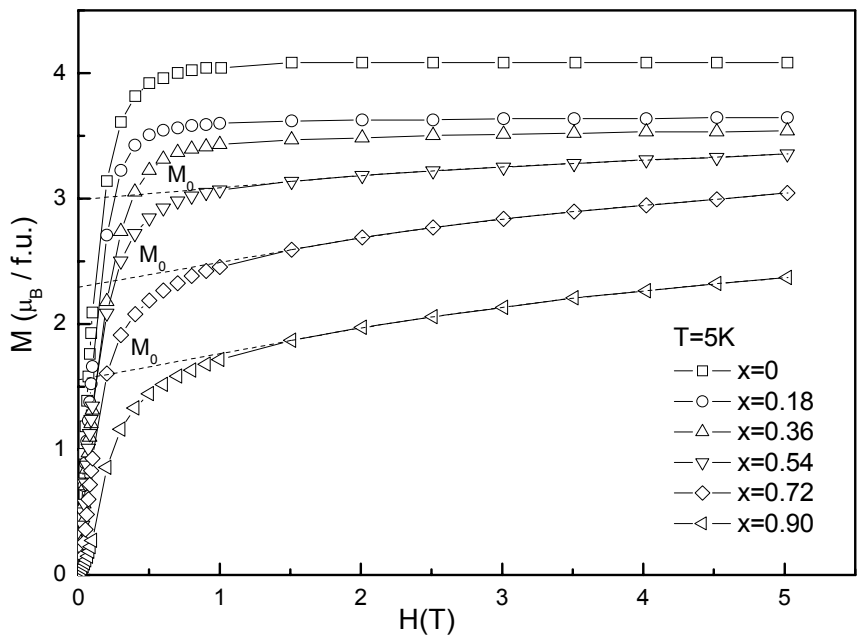

Fig.3 J. Yang et al.

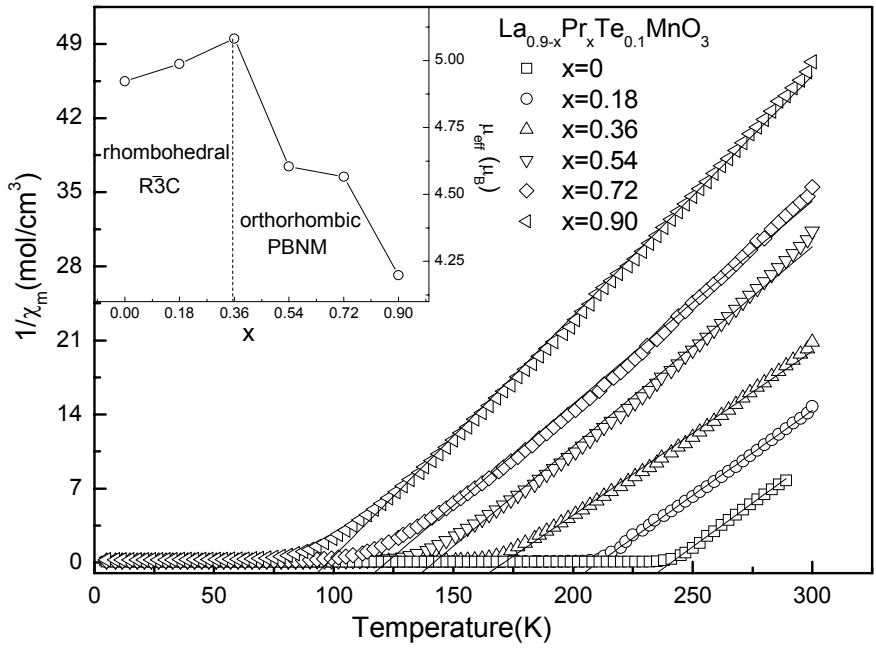

Fig.4 J. Yang et al.

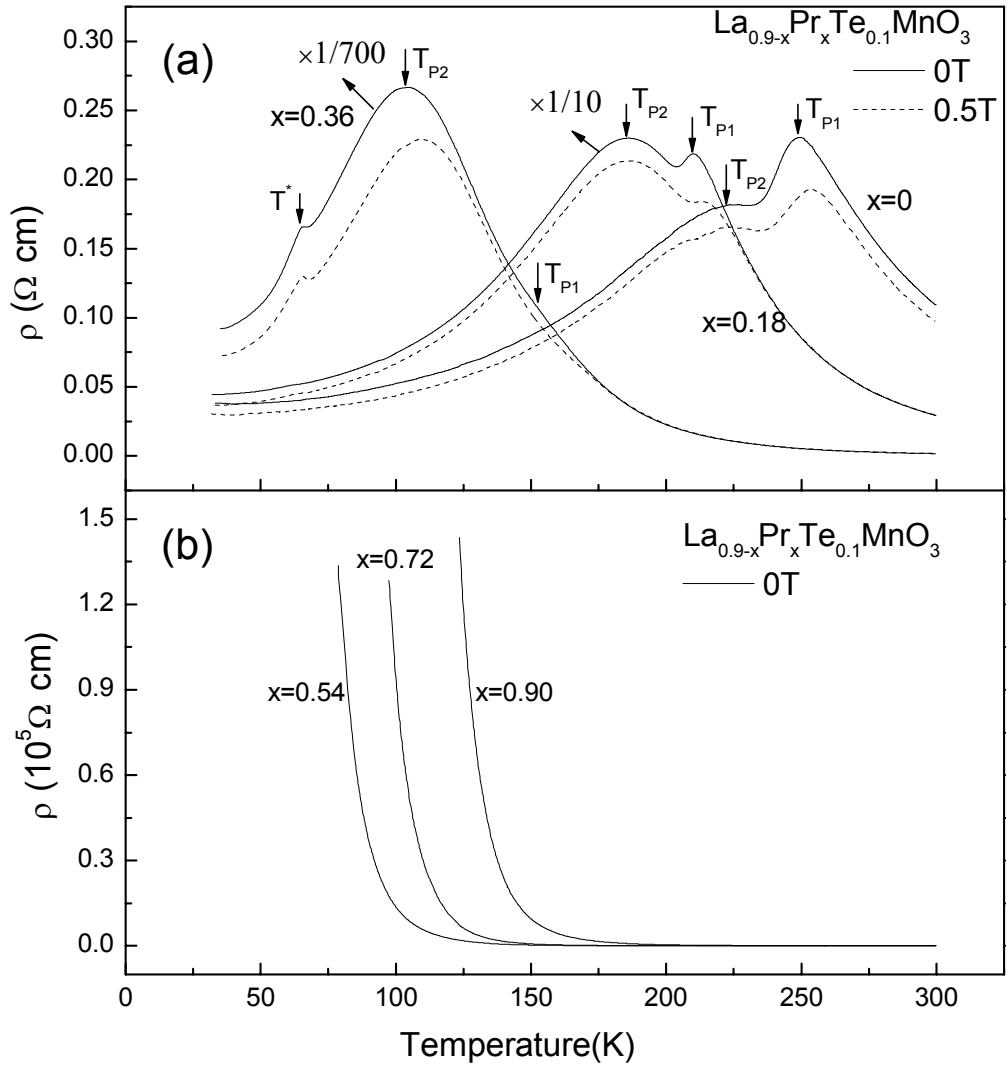

Fig.5 J. Yang et al.

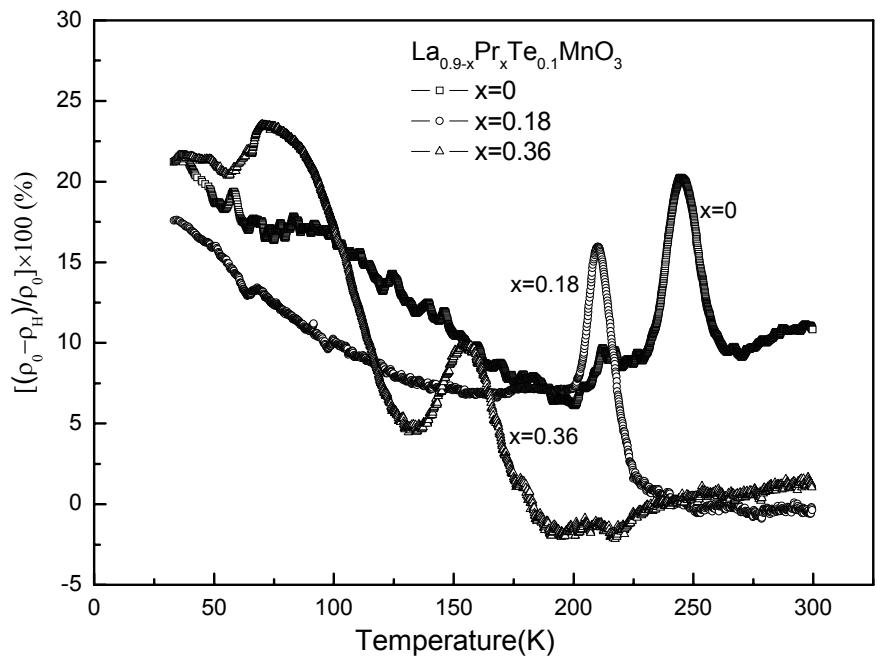

Fig.6 J. Yang et al.

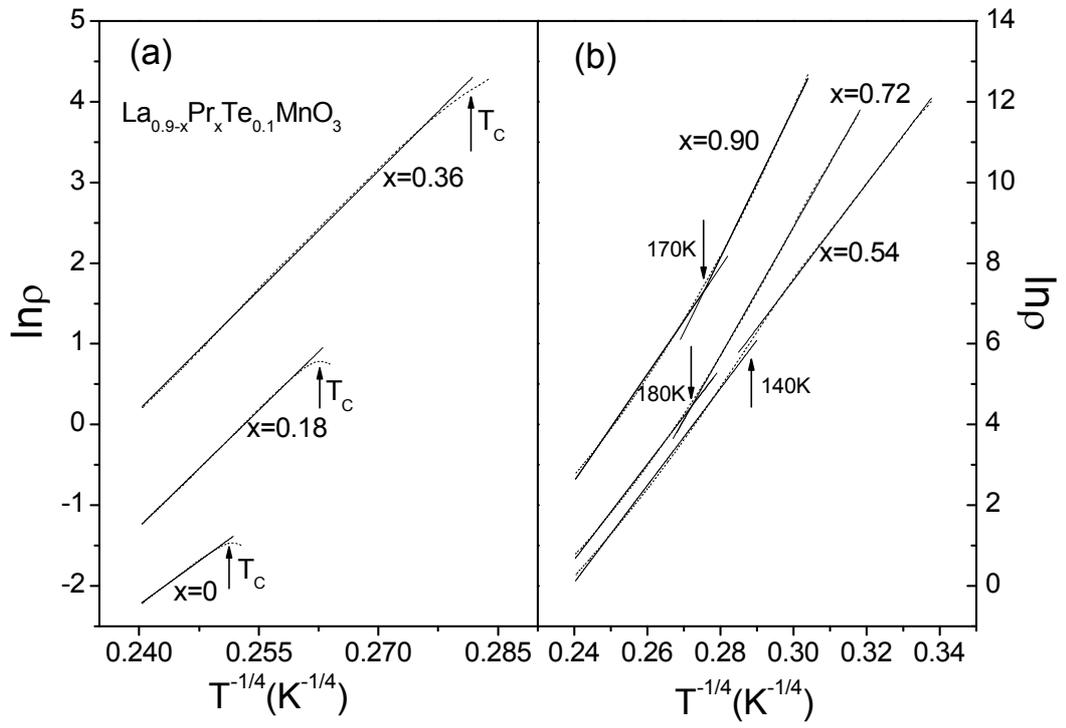

Fig.7. J. Yang et al.

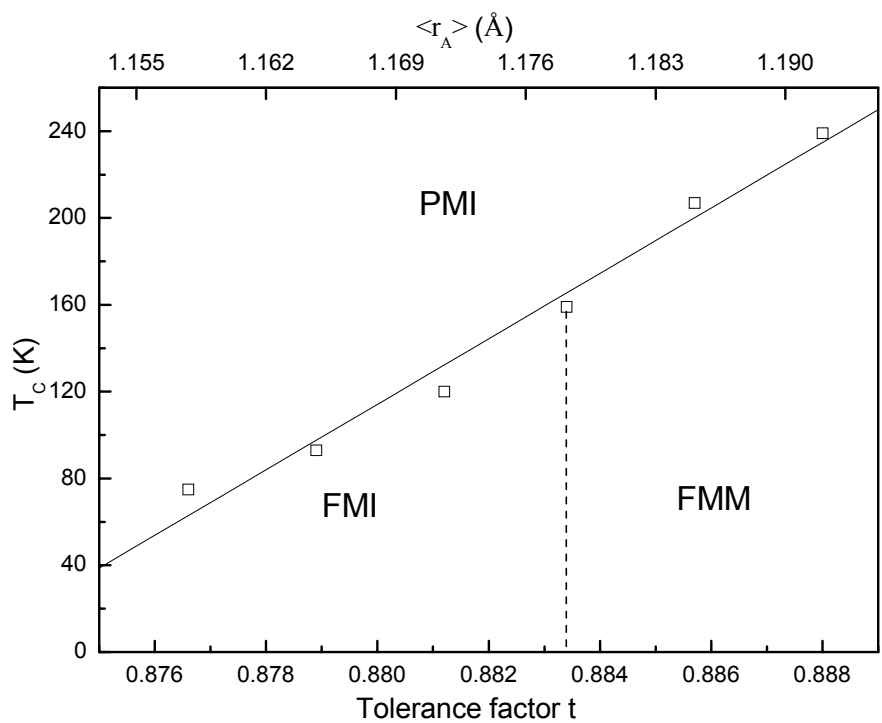

Fig.8 J. Yang et al.